\documentclass[conference]{IEEEtran}
\IEEEoverridecommandlockouts
\usepackage{cite}
\usepackage{amsmath,amssymb,amsfonts}
\usepackage{algorithm,algorithmic}
\usepackage{graphicx}
\usepackage{textcomp}
\usepackage{color,xcolor}
\usepackage{bm}
\usepackage{multirow}
\usepackage[normalem]{ulem}
\usepackage{tabularx}
\usepackage{cellspace}
\usepackage{xpatch}
\usepackage{soul}
\usepackage{ulem}
\usepackage{verbatim}
\usepackage{caption}
\usepackage{subcaption}

\usepackage[top=0.65in, bottom=1.22in, left=0.61in, right=0.62in]{geometry}

\def\BibTeX{{\rm B\kern-.05em{\sc i\kern-.025em b}\kern-.08em
    T\kern-.1667em\lower.7ex\hbox{E}\kern-.125emX}}
\begin{document}

\title{
	

MIMO Precoding Exploiting Extra Degrees of Freedom (DoF) in the Wavenumber Domain


}

\author{
{\large Yuanbin~Chen\textsuperscript{1},~Xufeng~Guo\textsuperscript{1},~Tianqi~Mao\textsuperscript{2,3},~Qingqing~Wu\textsuperscript{4},~Zhaocheng~Wang\textsuperscript{5},~Chau~Yuen\textsuperscript{1}}\\

{\normalsize \textsuperscript{1}School of Electrical and Electronic Engineering, Nanyang Technological University, Singapore 639798}\\
{\normalsize \textsuperscript{2}State Key Laboratory of CNS/ATM, Beijing Institute of Technology, Beijing 100081, China}\\
{\normalsize \textsuperscript{3}MIIT Key Laboratory of Complex-Field Intelligent Sensing, Beijing Institute of Technology, Beijing 100081, China }\\
{\normalsize \textsuperscript{4}Department of Electronic Engineering, Shanghai Jiao Tong University, Shanghai 200240, China}\\
{\normalsize \textsuperscript{5}Department of Electronic Engineering, Tsinghua University, Beijing 100084, China}


}

\maketitle

\begin{abstract}
In this paper, we propose an emerging wavenumber-domain precoding scheme to break the limitations of rank-1 channels that merely supports single-stream transmission, enabling simultaneous transmission of multiple data streams. The proposed wavenumber-domain precoding scheme also breaks the Rayleigh distance demarcation, regardless of the far-field and near-field contexts. Specifically, by characterizing the channel response as the superposition of a series of Fourier harmonics specified by different wavenumbers, the degree of freedom (DoF) is dependent on the cardinality of the wavenumber support, based on which the extra DoF is presented. This representation is applicable for both far-field and near-field. Different wavenumber atoms, determined within this support, constitute the codebook for MIMO precoding, in which each atom allows for the transmission of a data stream. Then, to maximize the capacity, it is required to select the wavenumbers associated with the optimal transmission direction, and optimize its power allocation. Finally, our simulation results demonstrate the significant superiority in comparison to the conventional spatial division schemes, with the potential of approaching the theoretical performance upper bound achieved by singular value decomposition (SVD).

\end{abstract}


\section{Introduction}

The footprint of multiple-input multiple-output (MIMO) systems is advancing towards larger array aperture sizes and higher frequency bands, inevitably resulting in the emergence of near-field communications \cite{XLM-105-CST,chen-jsac3}. Unlike the plane-wave assumption made in the conventional far-field scenario, electromagnetic (EM) waves propagate as spherical waves within the near-field region demarcated by the Rayleigh distance. To navigate this paradigm shift, a great deal of existing studies on near-field contexts employ the Fresnel approximation of spherical waves, resulting in the polar-domain design \cite{XLM-1}. This is particularly evident in the realms of channel state information (CSI) acquisition, such as channel estimation~\cite{guo-tvt}, beam training~\cite{wch-tvt}, and codebook design~\cite{XLM-16-CS-TWC}. When it comes to the precoding design, as the communication distance transitions into the near-field region, it becomes feasible to generate radiation patterns that concentrate the beam on a specific location, 
in which some efforts have been devoted to achieving beam focusing within the near field~\cite{XLM-7}. 

Regarding the polar-domain regime discussed in \cite{XLM-1,guo-tvt,wch-tvt}, we deem that it is a natural extension of the angular domain, whose essence lies in the uniform sampling of angles, similar to the angular domain, complemented by non-uniform sampling of the distance parameter at specified angular directions. 
Such an exhaustive sampling approach, even under the angle-distance de-correlation design criteria mentioned in~\cite{XLM-1,guo-tvt}, leads to an excessively large-size codebook, resulting in substantial computational complexity.
Secondly, designs based on the polar domain are only applicable for near-field scenarios. However, in practical situations, scatterers/UEs are typically located in both far-field and near-field regions, and priori on these locations is difficult to obtain. For example, when employing the multi-stage beam training presented in \cite{XLM-16-CS-TVT}, the use of the discrete Fourier transform (DFT) codebook for channel detection results in severe power leakage and entails additional pilots. Furthermore, other designs based on hybrid far-field and near-field scenarios rely on the Rayleigh distance, necessitating the use of tailored gearboxes for distinctively treating spherical waves and plane waves, respectively~\cite{subarray1}.

Inspired by the wavenumber-domain representation, which originates from describing the EM wave propagation in the sense of arbitrary scattering characteristics of the EM field, some fledging efforts have been devoted to employing this methodology to achieve a unified channel representation~\cite{chen-cm}. 
In our prior work~\cite{chen-cm}, we refined the wavenumber-domain representation discussed in \cite{Holo-2} by incorporating the contributions of both evanescent and propagating waves. This refinement allows for the characterization of communication channels over any distances and under any scattering conditions, regardless of the near-field and far-field.

At present, there is scant research on the methodologies for carrying out precoding in the wavenumber domain. 
The motivation for  precoding in this domain is to enhance spatial multiplexing capabilities, thereby increasing system capacity. 
Upon examining traditional precoding techniques, it becomes evident that they mostly include doing singular value decomposition (SVD) on the MIMO channel in order to achieve spatial multiplexing. Hence, the greatest level of precoding performance is dictated by the rank of the channel achieved by SVD, which indicates the largest number of data streams that can be allowed, also referred to as the degree of freedom (DoF)~\cite{Holo-35}. For the simplest single-user single-input single-output (SISO) case, the DoF is typically one. For practical MIMO implementations, the number of allowed data streams is typically limited to the number of RF chains, which is one of the critical performance bottlenecks of hybrid precoding architecture~\cite{hybrid_precoding}.
In response, we seek to transcend this limitation by exploiting the extra DoF available from both the propagating and evanescent waves at any given distance within the wavenumber domain.
This allows the channel response to be characterized as the superposition of a series of Fourier harmonics indexed by different wavenumbers. In this case, the DoF is determined by the cardinality of the wavenumber support. This makes it possible for the transmission of multiple data streams, even though over a rank-1 channel in the spatial domain, thus significantly enhancing the capacity of MIMO systems.

This work takes the first attempt at precoding in the wavenumber domain, aiming for filling a critical gap in this realm. Our contributions to advancing the state-of-the-art are summarized as follows:
\begin{itemize}
\item Given that the FPW approximation in the wavenumber domain is applicable over arbitrary scattering distance, we devise a novel precoding scheme for the extremely large-scale (XL) MIMO system of interest by leveraging the Fourier harmonic-based atoms. These wavenumber-domain atoms constitute the wavenumber-domain codebook, which breaks the Rayleigh distance limit and functions well in both far-field and near-field. 

\item The DoF of the wavenumber-domain codebook is determined by the  cardinality of the wavenumber support. Even for a rank-1 channel observed in the spatial domain, the extra DoF allows for the multiple data streams to be carried by the atoms (codewords) associated with specified wavenumbers, rather than a single data stream. Given the inter-wavenumber interference (IWI) caused by the limited directivity of the array, it is required to select the wavenumbers corresponding to the optimal transmission directions and allocate power accordingly directions to maximize capacity.

\item Simulation results are provided to evaluate the proposed precoding scheme in the wavenumber domain. Our results demonstrate that the proposed wavenumber-domain precoding scheme achieves capacity comparable to existing far-field counterparts. In the near-field, it fully harnesses additional DoF to significantly outperform the capacity performance of the spatial division scheme, approaching the performance upper bound achieved by SVD.

\end{itemize}

\section{System Model}\label{sec2}
\begin{figure}
	\centering
	\includegraphics[width=0.48\textwidth]{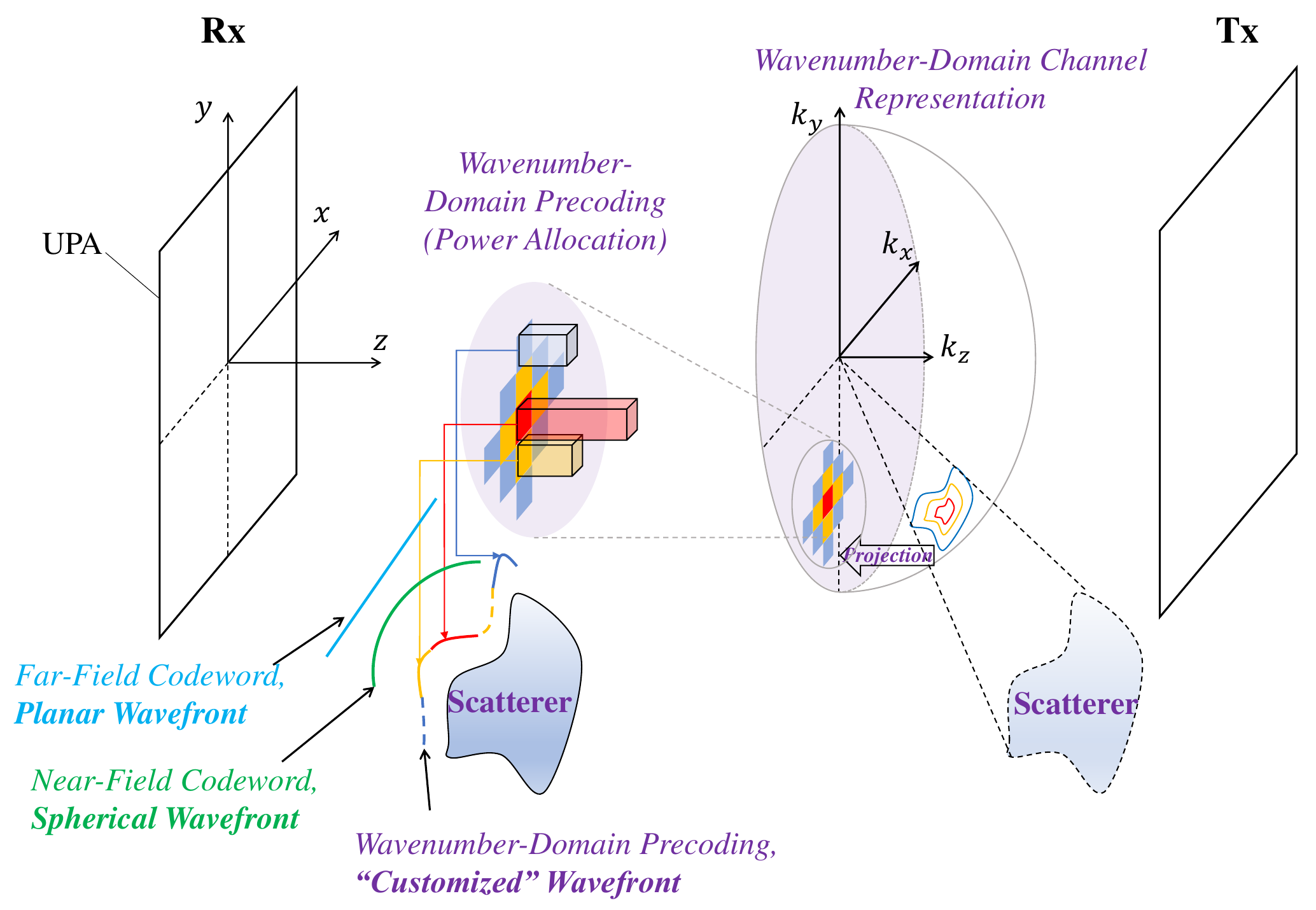}
	\caption{Illustration of the wavenumber-domain precoding.} \label{model}
 \vspace{-0.5cm}
\end{figure}

We consider a narrow-band XL-MIMO system with both the Tx and Rx equipped with uniform planar arrays (UPAs), as illustrated in Fig.~\ref{model}. The carrier frequency is  $f_c$ and the wavelength is $\lambda$. The size of the UPA at the Tx side is denoted by $L_T = L_{T,x} \times L_{T,y}$, while at the Rx it is  $L_R = L_{R,x} \times L_{R,y}$. The antenna spacing for both UPAs is uniformly maintained at half the wavelength, i.e., $\delta=\lambda/2$. Consequently, the number of antenna elements is explicitly given by $N_T = N_{T,x}N_{T,y}=L_{T}\delta^{-2}$ for the Tx and $N_R = N_{R,x}N_{R,y}=L_{R}\delta^{-2}$ for the Rx. The hybrid precoding architecture is adopted in the XL-MIMO system, where, in particular, the Tx is equipped with $N^{(T)}_{\text{RF}}$ radio frequency (RF) chains, while the Rx is equipped with $N^{(R)}_{\text{RF}}$ RF chains for the baseband signal processing. 
Hence, the transmit signal $\mathbf{x} \in \mathbb{C}^{N_T \times 1}$ can be given by $\mathbf{x} = \mathbf{F}_{\text{A}}\mathbf{F}_{\text{D}}\mathbf{s}$, where $\mathbf{s} \in \mathbb{C}^{N_{\text{RF}}^{(T)} \times 1}$, satisfying $\mathbf{s}^H\mathbf{s} = 1$, indicates the symbol vector fed into the RF chains.
Regarding the analog beamforming matrix $\mathbf{F}_{\text{A}}\in\mathbb{C}^{N_T\times N_{\text{RF}}^{(T)}}$, it can be achieved through the use of analog phase shifters.
Each entry of $\mathbf{F}_{\text{A}}$ is constrained to an equal-norm, given by $\left| \left[\mathbf{F}_{\text{A}}\right]_{i,j}\right|  = 1/\sqrt{N_T}$. 
This equal-norm constraint is indispensable for the hybrid architecture, since the practical phase shifter adjust the phase rather than the amplitude.
The digital beamforming matrix $\mathbf{F}_\text{D}$, denoted by $\mathbf{F}_\text{D} \in \mathbb{C}^{N_{\text{RF}}^{(T)}\times N_{\text{RF}}^{(T)}}$, complies with the power constraint  $\left\|  \mathbf{F}_\text{A}\mathbf{F}_\text{D}\right\| _F^2\leq P_T$. 
Then, by employing, at the Rx side, the combining matrix $\mathbf{C}_{\text{A}}\in\mathbb{C}^{N_{\text{RF}}^{(R)}\times N_R}$ that is equipped with analogy phase shifters confined with an equal-norm (i.e., $\left| {{{\left[ {{{\mathbf{C}}_{\text{A}}}} \right]}_{i,j}}} \right| = 1/\sqrt {{N_R}} $), 
the received signal $\mathbf{y}\in\mathbb{C}^{N_{\rm RF}^{(R)}\times 1}$ at the Rx can be given by
\begin{equation}\label{received_signal}
	\mathbf{y}=\mathbf{C}_\text{A}\mathbf{H}\mathbf{F}_\text{A}\mathbf{F}_\text{D}\mathbf{s}+\mathbf{n},
\end{equation}
where $\mathbf{n}\sim\mathcal{CN}(0,\sigma_0^2\mathbf{I}_{N^{(R)}_\text{RF}})$ denotes the additive white Gaussian noise (AWGN) and $\mathbf{H}\in\mathbb{C}^{N_R\times N_T}$ represents the channel matrix spanning from the Tx to the Rx.

\section{Wavenumber-Domain DoF Enhanced Precoding}
To demonstrate the extra DoF available in the wavenumber domain for the enhanced precoding, we commence by briefly reviewing the DoF obtained by the SVD technique in the conventional MIMO system, and then discuss the new DoF introduced in the wavenumber domain.

\subsection{DoF Achieved by the SVD Approach}
Let $\mathbf{H}=\mathbf{UDV} ^H$ be the SVD of the channel matrix, where $\mathbf{U}$ and $\mathbf{V}$ are the unitary matrices, and $\mathbf{D}$ is the diagonal matrix, denoted by ${\rm diag}\left( \sqrt{\lambda_1},\sqrt{\lambda_2},\dots,\sqrt{\lambda_{{\rm rank}(\mathbf{H})}},0,\dots \right)  \in \mathbb{C} ^{N_R\times N_T}$, with $\lambda_i$ representing the $i$-th singular value of $\mathbf{HH}^H$. Then, the channel matrix can be structured as
\begin{equation}\label{svdexpansion}
	\mathbf{H}=\sum_{i=1}^{{\rm rank}(\mathbf{H})}\sqrt{\lambda_{i}}\ \left[ \mathbf{U}\right] _{:,i}\left[ \mathbf{V}\right] ^H_{i,:},
\end{equation}
where $\left[ \mathbf{U}\right] _{:,i}$ denotes the $i$-th column of $\mathbf{U}$, and $\left[ \mathbf{V}\right] ^H_{i,:}$ represents the conjugated transpose of the $i$-th column of ${\bf V}$.
We suppose ${\rm rank}\left( \mathbf{H} \right) \leq\min\{N^{(R)}_{\text{RF}},N^{(T)}_{\text{RF}}\}$. 
To facilitate the subsequent theoretical analysis of channel capacity and DoF, we assume here that the amplitude and phase of the signal can be flexibly controlled by each antenna element{\footnote{This implies a fully digital XL-MIMO conception, which is impractical for implementation. The subsequent theoretical analysis provides a best-case bound. In Sec.~\ref{sec_simulation}, our simulation results are based on a more practical hybrid precoding architecture.}}. In this case, the MIMO channel can be independently decomposed into independent SISO channels with the number of ${\rm rank}\left( \mathbf{H}\right) $, by applying the precoding vector  $\left[ \mathbf{V}\right] _{:,i}$ and combining vector $\left[ \mathbf{U}\right] ^H_{:,i}$ for the $i$-th data stream.
Then, the capacity achieved by these independent SISO channels can be calculated by
\begin{equation}\label{capsvd}
	C_{\text{Fully-Digital}}=\sum^{{\rm rank}\left( \mathbf{H}\right) }_{i=1}\log \left(  1+\frac{p_i\lambda_i}{\sigma^2_0}\right) .
\end{equation}
where $p_i$ is the transmit power allocated to the $i$-th data stream, and $\left|  s_i \right| =1$ is the transmit symbol.
As intuitively observed in (\ref{capsvd}), the channel capacity of an XL-MIMO system is essentially the aggregation of the independent Shannon capacities of a series of SISO channels, with their number corresponding to the rank of $\mathbf{H}$, i.e., ${\rm{rank}}(\mathbf{H})$, also known as the DoF. Such kind of technique for capacity enhancement is referred to as the \textit{spatial multiplexing}.

In addition to the spatial multiplexing, the XL-MIMO system can also work in another mode, namely \textit{spatial division}. Specifically,
a number of data streams are fed into several SISO channels whose number is ${\rm rank}(\mathbf{H})$, and the channel capacity can be calculated by
\begin{equation}\label{capsd}
	C_{\text{Spatial-Division}}=\log\Big{(}1+\frac{\sum^{{\rm rank}(\mathbf{H})}_{i=1}p_i\lambda_i}{\sigma^2_0}\Big{)}.
\end{equation}
In examination of both (\ref{capsvd}) and (\ref{capsd}), it is evident that $C_{\text{Fully-Digital}} \geq C_{\text{Spatial-Division}}$ consistently holds true due to Jensen's inequality.
This implies that several orthogonal (low-SNR) SISO channels can provide a higher capacity compared to a single (high-SNR) SISO channel. More precisely, the spatial division focuses on enhancing the quality of a specific data stream to be transmitted by improving the SNR of that stream. By contrast, the spatial multiplexing concentrates on increasing the number of separate data streams that can be transmitted simultaneously. 
Hence, it is concluded that the terminology \textit{DoF} refers to the maximum number of orthogonal channels capable of transmitting independent information in the sense of spatial multiplexing~\cite{Holo-35}, denoted by $N_{\rm DoF}={\rm rank}(\mathbf{H})$. 
It becomes evident that when $N_{\rm DoF}>1$, we may have $C_{\text{Fully-Digital}} > C_{\text{Spatial-Division}} $.
Hence, the utilization of more DoF in spatial multiplexing results in a higher capacity compared to the spatial division.
Nevertheless, the codewords used in the Fully-Digital MIMO system fail to satisfy the equal norm constraint, i.e., $\left| {\left[ {\mathbf{V}} \right]_{i,j}^H} \right| \ne 1$ and $\left| {{{\left[ {\mathbf{U}} \right]}_{i,j}}} \right| \ne 1$. 
That is to say, in XL-MIMO systems, each RF chain feeds a information-carrying signal with the equal norm for each antenna element, in which the codewords, e.g., $\left[\mathbf{U}\right]_{:,i}$ and $\left[\mathbf{V}^H\right]_{i,:}$,  violate this constraint and thus cannot be applied in the hybrid precoding architecture. This is also a critical reason why the state-of-the-art MIMO designs focus on customized codebooks in the context of the conventional far-field circumstance.

Yet, when we shift our attention to the near-field context, these codebooks, based on the far-field assumption, become inapplicable. This discrepancy  primarily arises from the fundamental mismatch between the spherical EM waves and the plane-wave assumption. Imposing the classical DFT codebook on the near-field channel may result in a severe energy spread effect, significantly deteriorating the system performance~\cite{guo-tvt}. Fortunately, if we adopt a wavenumber domain perspective rather than an angular domain view for the precoding design, these codebooks are no longer dependent on the distance between scatterers and the Tx/Rx. Even more promising is that the extra DoF introduced by evanescent waves in the wavenumber domain has the potential to significantly enhance the capacity performance in the XL-MIMO system. Next, we transition the current channel into the wavenumber-domain perspective, in order to investigate new DoF to achive precoding enhancment.

\begin{figure*}[htbp]
	\begin{equation}\label{H_a}
		\small
		\left[{\bf H}\right]_{n_R, n_T} = 
		\sum_{\left(l_x, l_y\right) \in \xi_R}
		\sum_{\left(m_x, m_y\right) \in \xi_T}
		H_a\left(l_x,l_y,m_x,m_y\right) 
		\exp \left\lbrace j\left( \frac{2\pi l_xn_{R,x}\delta}{L_{R,x}}+\frac{2\pi l_yn_{R,y}\delta}{L_{R,y}}-\frac{2\pi m_xn_{S,x}\delta}{L_{S,x}}-\frac{2\pi m_yn_{S,y}\delta}{L_{S,y}}\right) \right\rbrace ,
	\end{equation}
	\hrule
\end{figure*}

\subsection{Wavenumber-Domain Channel Characterization}
By leveraging the wavenumber-domain representation~\cite{gyq-wcl,chen-cm}, the channel $\mathbf{H}$, detailed in (\ref{received_signal}), can be recast to  the superposition of a series of Fourier harmonics indexed by the wavenumbers, as shown in (\ref{H_a}) at the top of the next page. In (\ref{H_a}), $\xi_R$ and $\xi_T$ denote the wavenumber support associated with the Rx and the Tx, respectively, which can be specified as
\begin{subequations}\label{xi}
\begin{align}
		\xi_R&=\{(l_x,l_y)\in {\mathbb{Z}}^2:(\frac{2\pi l_x}{L_{R,x}})^2+(\frac{2\pi l_y}{L_{R,y}})^2\leq \beta k^2 \},
		\\
		\xi_T&=\{(m_x,m_y)\in {\mathbb{Z}}^2:(\frac{2\pi m_x}{L_{S,x}})^2+(\frac{2\pi m_y}{L_{S,y}})^2\leq \beta k^2 \}.
\end{align}
\end{subequations}
In (\ref{xi}), $\left( l_x, l_y\right) $ and $\left( m_x,m_y\right) $ represent the wavenumber indices associated with the Rx and the Tx, respectively. The parameter $\beta$ serves to retain the EM field magnitude excited by the evanescent waves at a significant level. Furthermore, we structure the channel in (\ref{H_a}) as its factorized matrix counterpart, i.e.,
\begin{equation}\label{wavenumberdomain}
		\mathbf{H}^{\text{WD}}=\mathbf{\Psi}_R\mathbf{H}_a{\mathbf{\Psi}}^H_T,
\end{equation}
where $\mathbf{\Psi}_R\in\mathbb{C}^{N_R\times \vert\xi_R\vert}$ and ${\mathbf{\Psi}}_T\in\mathbb{C}^{N_T\times \vert\xi_T\vert}$ are the wavenumber-domain  dictionary matrices at the Rx and Tx side, respectively, and the superscript ``WD" takes a shorthand for ``Wavenumber Domain". Upon denoting  $l\triangleq (l_x,l_y)\in\xi_R$, $m\triangleq (m_x,m_y)\in\xi_T$, the $\left( n_R,l\right) $-th entry of $ \mathbf{\Psi}_R $ and the $\left( {n_T,m}\right) $-entry of $ \mathbf{\Psi}_T $ are explicitly given by
\begin{subequations}\label{Psi}
	\begin{align}
		&[\mathbf{\Psi}_R]_{n_R,l}=\frac{1}{\sqrt{N_R}}\exp\Big\{j\big(\frac{2\pi l_xn_{R,x}\delta}{L_{R,x}}+\frac{2\pi l_yn_{R,y}\delta}{L_{R,y}}\big)\Big\}, 
  \\
		&[\mathbf{\Psi}_T]_{n_T,m}=\frac{1}{\sqrt{N_T}}\exp\Big\{j\big(\frac{2\pi m_xn_{T,x}\delta}{L_{T,x}}+\frac{2\pi m_yn_{T,y}\delta}{L_{T,y}}\big)\Big\}.
	\end{align}
\end{subequations}
Then, similar to the SVD procedure presented in (\ref{svdexpansion}), the wavenumber-domain channel $\mathbf{H}$ can be recast its SVD counterpart 
\begin{equation}\label{proposedexpansion}
\mathbf{H}^{\text{WD}}=\sum_{i=1}^{\xi_R}\sum_{j=1}^{\xi_T} \left[ \mathbf{\Psi}_R\right] _{:,i}\left[ \mathbf{H}_a\right] _{i,j}\left[ \mathbf{\Psi}^H_T\right] _{j,:} \, ,
\end{equation} 
where $\left[ \mathbf{\Psi}_R\right] _{:,i}$ denotes the $i$-th column of $\mathbf{\Psi}_R$, and $[\mathbf{\Psi}^H_T]_{j,:}$ indicates the $j$-th row of $\mathbf{\Psi}^H_T$.

Note that there are three significant differences between the channel representation shown in (\ref{svdexpansion}) and (\ref{proposedexpansion}). More explicitly, firstly, $\left[ \mathbf{U}\right] _{:,i}$ and $\left[ \mathbf{V}\right] ^H_{j,:}$ in (\ref{svdexpansion}) may break the equal-norm criterion, while both $\left[ \mathbf{\Psi}_R\right] _{:,i}$ and $\left[ \mathbf{\Psi}_T\right] ^H_{j,:}$ in (\ref{proposedexpansion}) do satisfy the equal-norm condition, i.e., $\vert \mathbf{\Psi}_R (i,j)\vert=1/\sqrt{N_R}$ and $\vert \mathbf{\Psi}_T (i,j)\vert=1/\sqrt{N_R}$. This implies that it is practical to generate codebooks with our proposed wavenumber-domain extension for the hybrid precoding framework, while leveraging (\ref{svdexpansion}) for codebook design is rather challenging. Secondly, in (\ref{svdexpansion}), {\color{black}the $i$-th transmit spatial codeword, denoted by $\left[ \mathbf{V}\right] ^H_{i,:}$}, corresponds to the $i$-th received spatial codeword, represented by $\left[ \mathbf{U}\right] _{:,i}$. By contrast, in (\ref{proposedexpansion}), {\color{black}the $j$-th transmit spatial codeword, $[\mathbf{\Psi}_T]^H_{j,:}$,} may not align with the $i$-th received spatial codeword $[\mathbf{\Psi}_R]_{:,i}$. This phenomenon may be attributed to a mismatch between the transmit direction and the received direction associated with a spatial subchannel. 
To tackle this issue, a delicate selection of specific wavenumbers is required.
We call this phenomenon as the \textit{inter-wavenumber interference (IWI)}. The IWI can be effectively mitigated by exploiting customized power allocation algorithm, as will be elaborated on later.

\subsection{Capacity Enhanced by Extra DoF in the Wavenumber Domain }

The number of data streams allowed is the minimum cardinality of the wavenumber support at the Tx or the Rx side, i.e., $K=\min\{ \left|  \xi_R\right| , \left| \xi_T \right|  \}$, indicating that the Tx is able to send $K$ data streams. We thus call the minimum cardinality of the wavenumber support as the \textit{extra DoF}, in comparison to the DoF achieved in (\ref{svdexpansion}). Then, the vector $\mathbf{x}^{\text{WD}}$ consisting of $K$ data streams can be given by
\begin{equation}\label{x_WD}
	\mathbf{x}^{\text{WD}}=\sum_{k=1}^{K} \sqrt{p_k}\ \mathbf{V}^{\text{WD}}_k s_k.
\end{equation}
in which $\mathbf{V}^{\text {WD}}_k\in \mathbb{C}^{N_T\times1}$ represents the transmit spatial codeword for the $k$-th data stream, which is selected from the columns of $\mathbf{\Psi}_T$, i.e., $\mathbf{V}^{\text{WD}}_k=\left[ \mathbf{\Psi}_T\right]_{:,c_k}$, with $\left[ \mathbf{\Psi}_T\right]_{:,c_k}$ indicating the $c_k$-th column of $\mathbf{\Psi}_T$. 
To receive the $k$-th data stream, the received antenna array borrows the combiner $\mathbf{U}^{\text{WD}}_k\in\mathbb{C}^{1\times N_R}$ to  process the received symbol, yielding that 
\begin{equation}\label{ywdkscalar}
	\left[ \mathbf{y}^{\text{WD}}\right]_k=\sqrt{p_k}\left[ \mathbf{H}_a\right] _{r_k,c_k}s_k+\sum_{m=1,m\neq k}^{K} \sqrt{p_k} \left[ \mathbf{H}_a\right] _{r_k,c_m}s_m+n_k,
\end{equation}
where $\mathbf{U}^{\text{WD}}_k$  can be picked from the rows in the matrix $\mathbf{\Psi}_R^H$, i.e., $\mathbf{U}^{\text{WD}}_k=\left[ \mathbf{\Psi}_R^H\right]_{r_k,:}$, with $\left[ \mathbf{\Psi}_R^H\right]_{r_k,:}$ indicating the $r_k$-th row of the matrix $\mathbf{\Psi}_R^H$. 
Assuming that the decoding is carried out by treating the interference as Gaussian noise, the signal-to-interference-plus-noise (SINR) for the $k$-th data stream can be determined using the following calculation
\begin{equation}
	{\rm SINR}_k=\frac{\left| \left[ \mathbf{H}_a\right] _{r_k,c_k}\right| ^2p_k}{\sum_{m=1,m\neq k}^{K} \left| \left[ \mathbf{H}_a\right] _{r_k,c_m}\right| ^2p_m+\sigma_0^2}.
\end{equation}
The capacity of the XL-MIMO system adopting the proposed wavenumber-domain extension can thus be determined by the total capacities aggregated by all $k$ data streams, which is calculated by
\begin{equation}\label{cap}
	C^{\text{WD}}=\sum_{k=1}^{K}\log\left( 1+{\rm SINR}_k\right) .
\end{equation}
The maximization of the capacity in (\ref{cap}) is determined by the two-fold aspects. On the one hand, for the $k$-th data stream, the values of $c_k$ and $r_k$ determine the transmission direction of the data stream, i.e., the spatial codewords. Therefore, it is necessary to carefully select these wavenumbers that can maximize capacity, constituting the optimal spatial codewords. On the other hand, when coping with IWI, it is essential to properly conduct the power allocation for each data stream that corresponds to a particular wavenumber.



\subsection{Solution: Maximize the Capacity in (\ref{cap})}

Our goal thus far is to maximize capacity in the wavenumber domain, given in (\ref{cap}), involving two key steps: wavenumber selection and power allocation to the selected wavenumbers. The following subsections details each of these steps.

\subsubsection{Step 1: Wavenumber Selection}\label{wavenumberselection}

The goal of the wavenumber selection is to determine the index pair $(r_k,c_k)$. Specifically, 
this process requires identifying the matching relationship between the 
the received codewords $\left[\mathbf{\Psi}_R\right]_{:,i}$ and the transmit codewords, denoted by $\left[\mathbf{\Psi}^H_T\right]_{j,:}$. Let $\mathbf{M}\in\{0,1\}^{\left| \xi_R\right|\times\left|\xi_T\right|}$ denote the wavenumber-selection matrix. Each entry of the matrix $\mathbf{M}$, denoted by $\bar{m}_{n_R,n_T}=\left[\mathbf{M}\right]_{n_R,n_T}$, is a binary variable (i.e., $\bar{m}_{n_R,n_T}\in{0,1}$). Specifically, $\bar{m}_{n_R,n_T}=1$ indicates that the $n_R$-th received codeword $\left[\mathbf{\Psi}_R\right]_{:,n_R}$ corresponds to the $n_T$-th transmit codeword $\left[\mathbf{\Psi}^H_T\right]_{n_T,:}$, while $\bar{m}_{n_R,n_T}=0$ indicates no correspondence.
Then, the wavenumber selection can be formulated as the following problem
\begin{subequations}\label{problem1}\small
\begin{align}
\arg\max_{\mathbf{M}} \ \ &\sum_{n_R=1}^{\left|\xi_R\right|}\sum_{n_T=1}^{\left|\xi_T\right|}\left|[\mathbf{H}_a]_{n_R,n_T}\right|^2  \bar{m}_{n_R,n_T}
\\
{\rm s.t.}\ &\sum_{n_R=1}^{\left| \xi_R\right|} \bar{m}_{n_R,n_T}=1, \forall n_T\in\left\lbrace1,2,\dots,\left|\xi_T\right| \right\rbrace,
\\
&\sum_{n_T=1}^{\left|\xi_T\right|} \bar{m}_{n_R,n_T}=1, \forall n_R\in\left\lbrace 1,2,\dots,\left|\xi_R\right|\right\rbrace,
\\
&\bar{m}_{n_R,n_T}\in\{0,1\}.
\end{align}
\end{subequations} 
Problem (\ref{problem1}) a linear integer programming problem involving integer variables, and it can be efficiently solved by employing the `\texttt{intlinprog}' package in PYTHON.

\subsubsection{Step 2: Power Allocation to the Selected Wavenumbers}\label{powerallocation}
After obtaining the optimal wavenumbers associated with data stream $k$, the power allocated to this stream has to be determined. Let $\mathbf{p}=[p_1,p_2,\dots,p_{K}]^T$ denote the power allocated to each data stream. By appropriately allocating the power to each data stream associated with the specific wavenumber, the IWI can be effectively mitigated. Thus, the capacity-maximization problem can be formulated as
\begin{subequations}\label{problem2}\small
\begin{align}
\arg\max_{\mathbf{p}}\sum_{k=1}^{K} &\log\Big{(}1+\frac{\vert[\mathbf{H}_a]_{r_k,c_k}\vert^2p_{k}}{\sum_{m=1,m\neq k}^{K}\vert[\mathbf{H}_a]_{r_k,c_m}\vert^2p_{m}+\sigma^2_0}\Big{)}
\\
&{\rm s.t.}\ \sum_{k=1}^{K}p_{k}\leq P_T,\ p_k\geq0.
\end{align}
\end{subequations}
Problem (\ref{problem2}) is a non-convex problem due to the non-concavity in the objective function. To circumvent this issue, we introduce the slack variable $\mathbf{u}=\left[u_1,u_2,\dots,u_K\right]^T$, thus yielding the following problem 
\begin{subequations}\label{cvxproblem}\small
\begin{align}
\arg&\max_{\mathbf{p,u}} \ \sum_{k=1}^{K}  \log\Big{(}1+\frac{\vert[\mathbf{H}_a]_{r_k,c_k}\vert^2p_{k}}{u_k+\sigma^2_0}\Big{)}
\\
{\rm s.t.}\ &\sum_{k=1}^{K}p_{k}\leq P_T,\ p_k\geq0,
\\
&u_k\geq\sum_{m=1,m\neq k}^{K}\vert[\mathbf{H}_a]_{r_k,c_m}\vert^2p_{m},\forall k\in\{1,2,\dots,K\},
\end{align}
\end{subequations}
where the objective function is jointly concave with variables $\mathbf{p}$ and $\mathbf{u}$, and the constraints in (\ref{problem2}b)-(\ref{problem2}c) are convex. Therefore, problem (\ref{cvxproblem}) is a standard convex optimization program that can be solved using off-the-shelf convex solvers, such as CVX~\cite{cvx}. Note that Step 1 and Step 2 are independent of each other. The output from Step 1, i.e., the optimal wavenumber selected, is directly fed into Step 2 for power allocation, aiming for maximizing capacity. Therefore, the precoding scheme in the wavenumber domain does not require unnecessary iterations, significantly improving the computational efficiency.

\begin{figure*}[t]
	\centering
	\begin{subfigure}[b]{0.32\linewidth}
		\centering
		\includegraphics[width=0.98\linewidth]{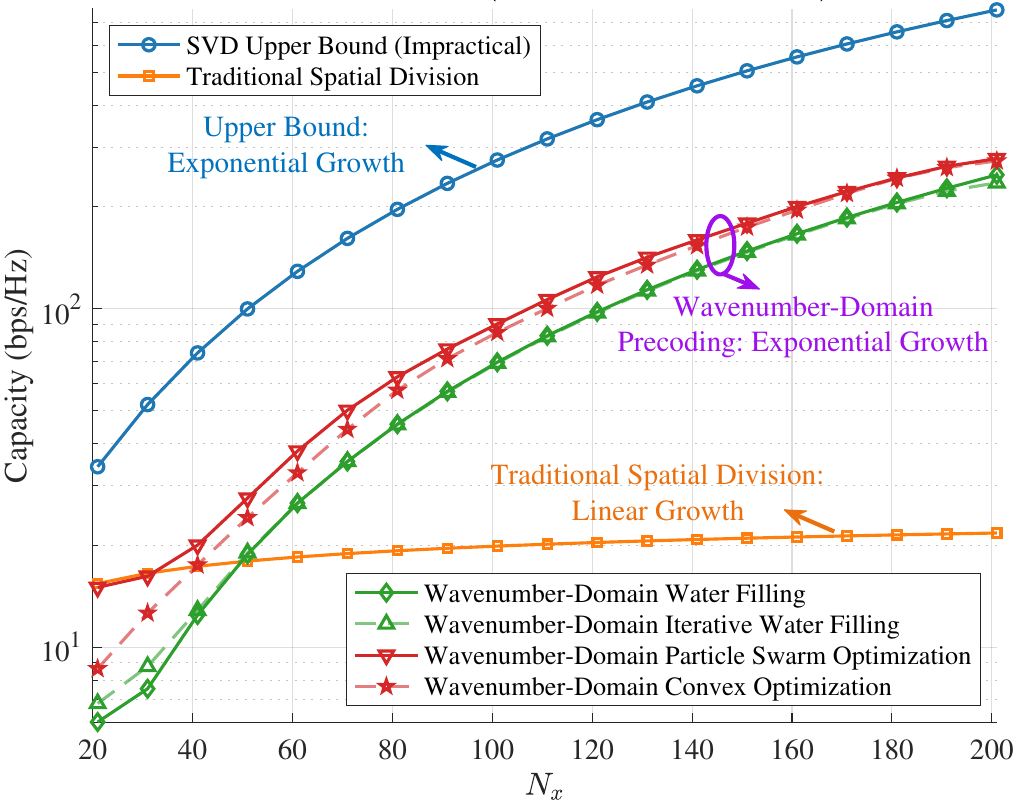}
		\caption{}
	\end{subfigure}
	\hfill
	\begin{subfigure}[b]{0.32\linewidth}
		\centering
		\includegraphics[width=0.98\linewidth]{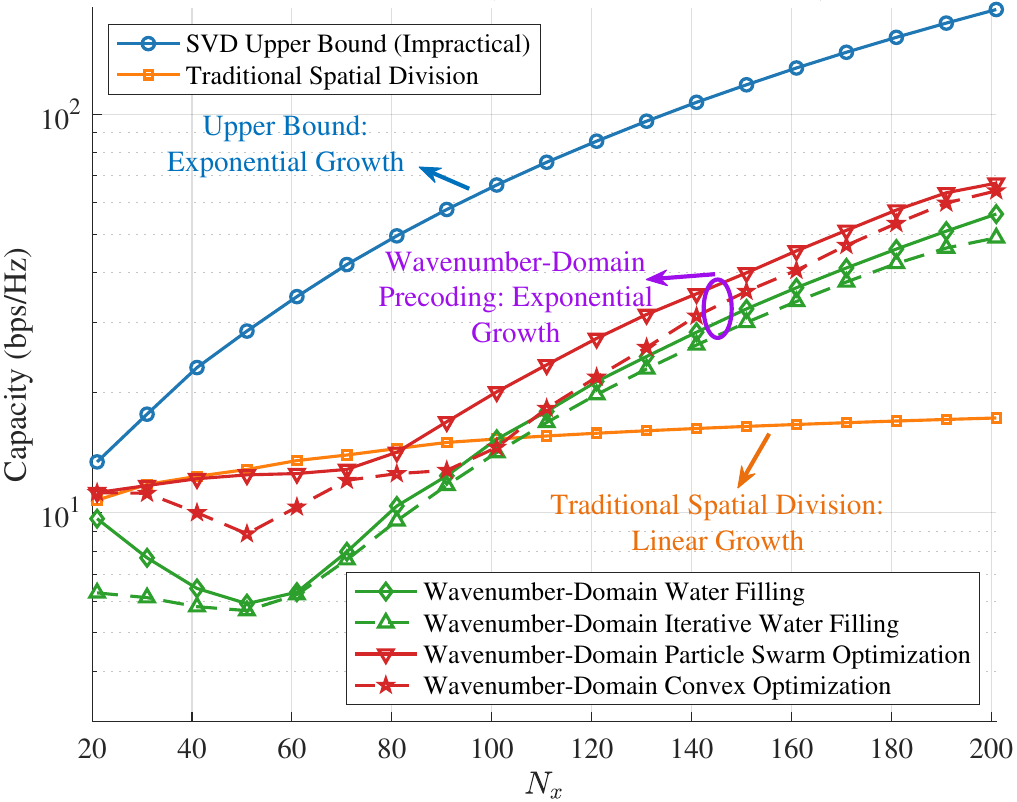}
		\caption{}
	\end{subfigure}
	\hfill
	\begin{subfigure}[b]{0.32\linewidth}
		\centering
		\includegraphics[width=0.98\linewidth]{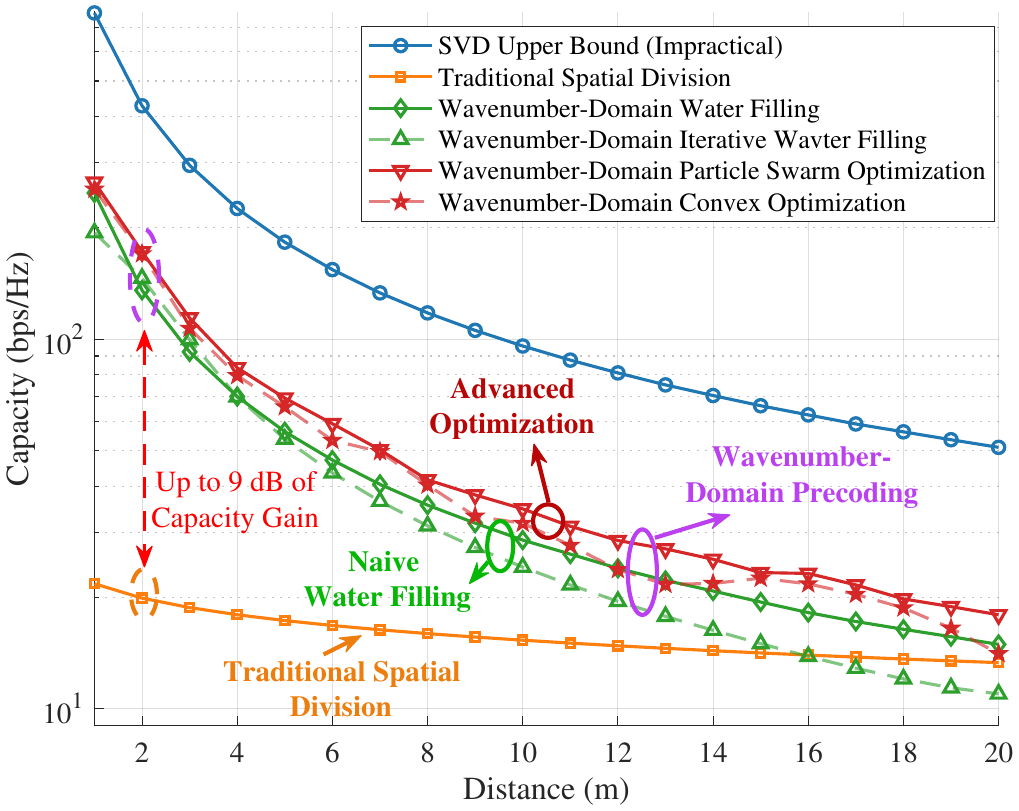}
		\caption{}
	\end{subfigure}
	\caption{(a) Capacity versus $N_x=N_{R,x} = N_{T,x}$ with Tx-Rx distance of $1$ meters. (b) Capacity versus $N_x=N_{R,x} = N_{T,x}$ with Tx-Rx distance of $5$ meters. (c) Capacity versus distance.}
	\label{fig:CapNx}
\end{figure*}

\section{Simulation Results}\label{sec_simulation}

This section demonstrates simulation results to evaluate our proposed wavenumber-domain precoding scheme in the  XL-MIMO system of interest. The number of antenna elements at the Tx and Rx are equipped in the form of $N_{T,x} = N_{R,x} = 129$ and $N_{T,y} = N_{R,y} = 9$. 
The considered XL-MIMO system operates at 30~GHz with a bandwidth of 300~MHz. It is assumed that the number of scatterers is $Q = 2$. Furthermore, the complex gain of the LoS path ($q=1$) is modeled to follow a complex Gaussian distribution $\mathcal{CN} \left( 0,1\right) $, and for the NLoS path ($q \ge 2$), the complex gain is characterized by $\mathcal{CN} \left( 0,0.01\right) $. The total transmit power at the Tx is $P_T = 23~\rm{dBm}$ and the noise power is $-89~\rm{dBm}$. Our performance metric is the channel capacity presented in~(\ref{cap}). For comparison, we consider several benchmarks as follows: (i) \textbf{SVD} scheme, providing the theoretical upper bound of the capacity; (ii) \textbf{Spatial Division} scheme presented in (\ref{capsd}); (iii) \textbf{Water Filling} algorithm for solving problem (\ref{cap}); (iv) \textbf{Iterative Water Filling} algorithm for problem solving (\ref{cap})~\cite{Iterative_WF}; (v) \textbf{Particle Swarm Optimization} algorithm for resolving problem~(\ref{problem2})~\cite{PSO}.




As illustrated in Fig.~\ref{fig:CapNx}(a), we present the capacity versus the number of antenna elements with a fixed distance of 1~meter between the Tx and Rx.
Upon observation, the capacity achieves by all schemes grows with the increase of $N_x$, though at different growth patterns. This growth is anticipated, stemming from the enhanced DoF provided by the increased number of antenna elements, which facilitates more independent sub-channels for data transmission. Futhermore, our proposed wavenumber-domain regime, using different algorithms, delivers significantly greater capacity compared to that attained by the classical spatial diversity. This enhancement is primarily due to the division of the original single data stream channel into multiple wavenumber domain subchannels, allowing parallel data transmission in the wavenumber domain.  This division also corresponds to the extra DoF inherent in the wavenumber-domain channel.

Fig.~\ref{fig:CapNx}(b) examines the capacity performance versus the number of antenna elements, $N_x = N_{R,x} = N_{T,x}$, at a Tx-Rx distance of $5~\rm{m}$. We observe two intriguing phenomena. Firstly, when $N_x < 80$, the capacities achieved by the wavenumber-domain based schemes are inferior to those obtained via the spatial division. Secondly, the capacity performance of the wavenumber-domain scheme initially decreases and then increases as $N_x$ increases. To delve into the reasons behind these observations, we conduct the following analysis.
Specifically, when the array aperture size is relatively small (e.g., $N = 20$), the beamwidth significantly exceeds the wavenumber resolution determined by the array aperture size (i.e., the interval between wavenumbers). Given that the wavenumber interval does not surpass the beamwidth, there is severe IWI, thus limiting the capacity achieved by the wavenumber-domain approach. As $N_x$ increases, the beamwidth gradually becomes wider, allowing more power to be allocated to larger wavenumbers to mitigate the wavenumber interference. This explains why the performance gap between the two kind of schemes grows significantly with the array aperture size.



In Fig.~\ref{fig:CapNx}(c), we demonstrate the capacity versus the Tx-Rx distance, where the number of antenna elements is $N_x=200, N_y=3$. As it transpires, the capacity achieved by all schemes decrease with the increased distance between the Tx and Rx.
This occurs because when the Tx and Rx are positioned at larger distances from each other, they behave like point sources in the far-field region. In this case, the number of DoF (determined by SVD) converges to 1, leading to a significant decrease in the theoretical channel capacity. In spite of such performance erosion, the capacities achieved by the wavenumber-domain schemes still exhibit superior capacity performance to the traditional spatial division schemes. 
For example, when the distance is small, the performance gap between the wavenumber-domain schemes and the spatial division may reach upwards of $9~\rm{dB}$ when the distance is $2~\rm{m}$.

\section{Conclusion}
By leveraging the FPW approximation in the wavenumber domain, this paper devises an XL-MIMO hybrid precoding codebook using Fourier-harmonic-based atoms, which is applicable for both far-field and near-field conditions. This approach also allows for the transmission of multiple data streams, even though over a rank-1 channel. Then, taking into account the IWI caused by limited array directivity, we propose an effective solution for selecting wavenumbers and allocating power. Finally, simulation results demonstrate that the proposed wavenumber-domain precoding scheme can achieve the same capacity performance as existing systems in the far-field. Furthermore, it utilizes the extra DoF available in the near field that significantly exceed the capacity performance of the spatial division.

\bibliographystyle{IEEEtran}
\bibliography{GCw_24}

\begin{thebibliography}{10}
\providecommand{\url}[1]{#1}
\csname url@samestyle\endcsname
\providecommand{\newblock}{\relax}
\providecommand{\bibinfo}[2]{#2}
\providecommand{\BIBentrySTDinterwordspacing}{\spaceskip=0pt\relax}
\providecommand{\BIBentryALTinterwordstretchfactor}{4}
\providecommand{\BIBentryALTinterwordspacing}{\spaceskip=\fontdimen2\font plus
\BIBentryALTinterwordstretchfactor\fontdimen3\font minus
  \fontdimen4\font\relax}
\providecommand{\BIBforeignlanguage}[2]{{%
\expandafter\ifx\csname l@#1\endcsname\relax
\typeout{** WARNING: IEEEtran.bst: No hyphenation pattern has been}%
\typeout{** loaded for the language `#1'. Using the pattern for}%
\typeout{** the default language instead.}%
\else
\language=\csname l@#1\endcsname
\fi
#2}}
\providecommand{\BIBdecl}{\relax}
\BIBdecl

\bibitem{XLM-105-CST}
H.~Lu, Y.~Zeng, C.~You, Y.~Han, J.~Zhang, Z.~Wang, Z.~Dong, S.~Jin, C.-X. Wang,
  T.~Jiang, X.~You, and R.~Zhang, ``A tutorial on near-field {XL-MIMO}
  communications towards {6G},'' \emph{IEEE Commun. Surv. Tutor.}, to appear,
  2024.

\bibitem{chen-jsac3}
Y.~Chen, Y.~Wang, Z.~Wang, and Z.~Han, ``Angular-distance based channel
  estimation for holographic {MIMO},'' \emph{IEEE J. Sel. Areas Commun.},
  vol.~42, no.~6, pp. 1684--1702, Jun. 2024.

\bibitem{XLM-1}
M.~Cui and L.~Dai, ``Channel estimation for extremely large-scale {MIMO}:
  Far-field or near-field?'' \emph{IEEE Trans. Commun.}, vol.~70, no.~4, pp.
  2663--2677, Apr. 2022.

\bibitem{guo-tvt}
X.~Guo, Y.~Chen, and Y.~Wang, ``Compressed channel estimation for near-field
  {XL-MIMO} using triple parametric decomposition,'' \emph{IEEE Trans. Veh.
  Technol.}, vol.~72, no.~11, pp. 15\,040--15\,045, Nov. 2023.

\bibitem{wch-tvt}
C.~Weng, X.~Guo, and Y.~Wang, ``Near-field beam training with hierarchical
  codebook: Two-stage learning-based approach,'' \emph{IEEE Trans. Veh.
  Technol.}, to appear, 2024.

\bibitem{XLM-16-CS-TWC}
X.~Wu, C.~You, J.~Li, and Y.~Zhang, ``Near-field beam training: Joint angle and
  range estimation with dft codebook,'' \emph{IEEE Trans. Wireless Commun.}, to
  appear, 2024.

\bibitem{XLM-7}
H.~Zhang, N.~Shlezinger, F.~Guidi, D.~Dardari, M.~F. Imani, and Y.~C. Eldar,
  ``Beam focusing for near-field multiuser {MIMO} communications,'' \emph{IEEE
  Trans. Wireless Commun.}, vol.~21, no.~9, pp. 7476--7490, Sep. 2022.

\bibitem{XLM-16-CS-TVT}
C.~Wu, C.~You, Y.~Liu, L.~Chen, and S.~Shi, ``Two-stage hierarchical beam
  training for near-field communications,'' \emph{IEEE Trans. Veh. Technol.},
  vol.~73, no.~2, pp. 2032--2044, Feb. 2024.

\bibitem{subarray1}
Y.~Chen, R.~Li, C.~Han, S.~Sun, and M.~Tao, ``Hybrid spherical- and planar-wave
  channel modeling and estimation for {Terahertz} integrated {UM-MIMO} and
  {IRS} systems,'' \emph{IEEE Trans. Wireless Commun.}, vol.~22, no.~12, pp.
  9746--9761, Dec. Dec. 2023.

\bibitem{chen-cm}
Y.~Chen, X.~Guo, G.~Zhou, S.~Jin, D.~W.~K. Ng, and Z.~Wang, ``Unified far-field
  and near-field in holographic {MIMO}: A wavenumber-domain perspective,''
  \emph{IEEE Commun. Mag.}, to appear, 2024, available online:
  \url{https://arxiv.org/abs/2407.14815}.

\bibitem{Holo-2}
A.~Pizzo, L.~Sanguinetti, and T.~L. Marzetta, ``Fourier plane-wave series
  expansion for holographic {MIMO} communications,'' \emph{IEEE Trans. Wireless
  Commun.}, vol.~21, no.~9, pp. 6890--6905, Sep. 2022.

\bibitem{Holo-35}
S.~S.~A. Yuan, Z.~He, X.~Chen, C.~Huang, and W.~E.~I. Sha, ``Electromagnetic
  effective degree of freedom of an {MIMO} system in free space,'' \emph{IEEE
  Antennas Wirel. Propag. Lett.}, vol.~21, no.~3, pp. 446--450, Mar. 2022.

\bibitem{hybrid_precoding}
X.~Yu, J.-C. Shen, J.~Zhang, and K.~B. Letaief, ``Alternating minimization
  algorithms for hybrid precoding in millimeter wave {MIMO} systems,''
  \emph{IEEE J. Sel. Topics Signal Process.}, vol.~10, no.~3, pp. 485--500,
  Apr. 2016.

\bibitem{gyq-wcl}
Y.~Guo, Y.~Chen, and Y.~Wang, ``Channel estimation for holographic {MIMO}:
  Wavenumber-domain sparsity inspired approaches,'' \emph{IEEE Wireless Commun.
  Lett.}, to appear, 2024.

\bibitem{cvx}
M.~Grant and S.~Boyd, ``{CVX}: Matlab software for disciplined convex
  programming, version 2.1,'' \url{https://cvxr.com/cvx}, Mar. 2014.

\bibitem{Iterative_WF}
G.~Scutari, D.~P. Palomar, and S.~Barbarossa, ``Asynchronous iterative
  water-filling for gaussian frequency-selective interference channels,''
  \emph{IEEE Trans. Inf. Theory}, vol.~54, no.~7, pp. 2868--2878, Jul. 2008.

\bibitem{PSO}
D.~Wang, D.~Tan, and L.~Liu, ``Particle swarm optimization algorithm: an
  overview,'' \emph{Soft computing}, vol.~22, no.~2, pp. 387--408, Jan. 2018.

\end{thebibliography}

\end{document}